\documentclass[aps, prl, reprint, showpacs, superscriptaddress]{revtex4-1}
\usepackage{amsmath, amssymb, graphicx, subfigure, CJK} 
\usepackage{mathptmx, color, hyperref}
\usepackage[all]{hypcap}

\newcommand{\figurescale}{1}

\definecolor{MyDarkGreen}{rgb}{0,0.6,0}
\definecolor{MyDarkBlue}{rgb}{0,0,0.8}
\definecolor{MyDarkRed}{rgb}{0.6,0,0.3}
\hypersetup{breaklinks=true,colorlinks=true,plainpages=true,linktocpage=true,linkcolor=MyDarkBlue,citecolor=MyDarkGreen,urlcolor=MyDarkRed,pdfborder={0 0 0},%
pdfauthor={Sang-Kil Son, Henry N. Chapman, and Robin Santra},%
pdftitle={Multi-wavelength anomalous diffraction at high x-ray intensity}%
}

\begin{document}
\begin{CJK*}{UTF8}{}

\title{Multi-wavelength anomalous diffraction at high x-ray intensity}


\author{Sang-Kil Son \CJKfamily{mj}(손상길)}
\email{sangkil.son@cfel.de}
\affiliation{Center for Free-Electron Laser Science, DESY, Hamburg, Germany}

\author{Henry N. Chapman}
\email{henry.chapman@cfel.de}
\affiliation{Center for Free-Electron Laser Science, DESY, Hamburg, Germany}
\affiliation{Department of Physics, University of Hamburg, Hamburg, Germany}

\author{Robin Santra}
\email{robin.santra@cfel.de}
\affiliation{Center for Free-Electron Laser Science, DESY, Hamburg, Germany}
\affiliation{Department of Physics, University of Hamburg, Hamburg, Germany}

\date{\today}

\begin{abstract}
The multi-wavelength anomalous diffraction (MAD) method is used to determine phase information in x-ray crystallography by employing dispersion corrections from heavy atoms on coherent x-ray scattering.
X-ray free-electron lasers (FELs) show promise for revealing the structure of single molecules or nanocrystals within femtoseconds, but the phase problem remains largely unsolved.
Due to the ultrabrightness of x-ray FEL, samples experience severe electronic radiation damage, especially to heavy atoms, which hinders direct implementation of the MAD method with x-ray FELs.
We propose a generalized version of the MAD phasing method at high x-ray intensity.
We demonstrate the existence of a Karle--Hendrickson-type equation for the MAD method in the high-intensity regime and calculate relevant coefficients with detailed electronic damage dynamics of heavy atoms.
Our results show that the bleaching effect on the scattering strength of the heavy atoms can be advantageous to the phasing method.
The present method offers a potential for \textit{ab initio} structural determination in femtosecond x-ray nanocrystallography.
\end{abstract}


\maketitle
\end{CJK*}

Determination of the 3D structure of proteins and macromolecules is crucial to understand their biological functions at the molecular level.
X-ray crystallography has been widely used for structural determination~\cite{Dauter06}, but it suffers from two bottlenecks: the phase problem and growing high-quality crystals.
The phase problem~\cite{Karle50,Taylor03} is a fundamental obstacle in constructing an electronic density map from x-ray diffraction.
Multi-wavelength anomalous diffraction (MAD)~\cite{Guss88,Hendrickson89,Hendrickson91} with synchrotron radiation is one of the major achievements to address this issue.
%
X-ray free-electron lasers (FELs)~\cite{McNeil10} promise to have a revolutionary impact on molecular imaging~\cite{Gaffney07,Neutze00}, overcoming the crystal bottleneck.
The unprecedented high x-ray fluence provides a sufficiently large number of photons to enable structure determination from diffraction measurements of streams of single molecules~\cite{Neutze00,Chapman06,Chapman10} and nanocrystals~\cite{Mancuso09,Chapman11}.
However, due to an extremely high fluence that is $\sim$100 times larger than the conventional damage limit~\cite{Henderson95}, samples are subject to severe radiation damage~\cite{Howells09}.
The ultrashort x-ray pulses generated by x-ray FELs enable us to carry out ``diffraction-before-destruction'' within femtosecond timescales to suppress nuclear motion~\cite{Neutze00}.
%
Nonetheless, electronic damage~\cite{Neutze00,Hau-Riege04,Quiney11} during femtosecond x-ray pulses is unavoidable, leading us to consider ``diffraction-during-ionization''~\cite{Son11a}.
This electronic radiation damage is particularly challenging when addressing the phase problem by anomalous dispersion~\cite{Ravelli05}, because heavy atoms as anomalous scatterers will be more ionized than other atoms during intense x-ray pulses.
%
Therefore, it has been speculated that MAD would not be an applicable route for phasing in the presence of severe radiation damage~\cite{Dauter06,Ravelli05}.
Here we propose a high-intensity version of the MAD phasing method based on a detailed description of the electronic response at the atomic level. 
In contrast to the speculation, our results show that MAD not only works, but also that the extensive electronic rearrangements at high x-ray intensity provide a new path to phasing.
We will then demonstrate that this approach is applicable to the phase problem in femtosecond x-ray nanocrystallography~\cite{Mancuso09,Chapman11}, which is one of the most prominent topics in x-ray FEL applications.

\begin{figure}
\includegraphics[scale=\figurescale]{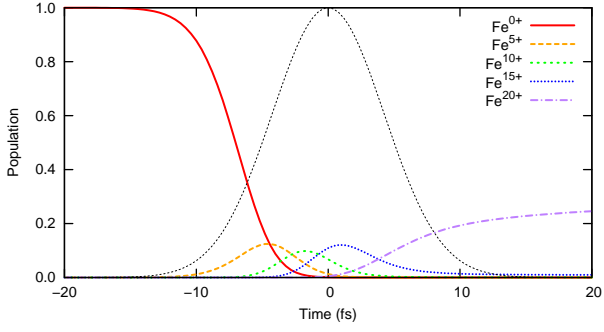}
\caption{(Color online) Population dynamics for several selected charge states of Fe during an x-ray pulse.
The thin dotted line indicates the Gaussian pulse envelope.  See the text for parameters used.}
\label{fig:population}
\end{figure}

X rays mainly ionize inner-shell electrons and subsequent relaxation (Auger decay and fluorescence) fills the inner-shell vacancy.
Therefore, sequences of photoionization and relaxation can strip off many electrons after absorbing several photons~\cite{Young10}.
For heavy atoms that have more than two subshells, a vacancy in deep inner-shells causes several relaxation steps in the cascade through the subshells, resulting in further electron ejections~\cite{Carlson66}.
To simulate the electronic damage dynamics, we use the \textsc{xatom} toolkit~\cite{Son11a,xatom}, where all rates and cross sections are calculated within the nonrelativistic Hartree--Fock--Slater method and multiphoton electronic dynamics is described by sequential one-photon processes with all possible electronic configurations.
Figure~\ref{fig:population} depicts the time evolution of populations for several charge states of an iron (Fe) atom where 27,783 coupled rate equations were solved.
The photon energy is 8~keV and the fluence is $5$$\times$$10^{12}~\text{photons}/\mu\text{m}^2$.
The pulse duration is 10~fs full-width-at-half-maximum (FWHM) with a Gaussian envelope.
In these conditions, the neutral Fe is completely depleted and high charge states such as Fe$^{20+}$ are substantially produced by the end of the pulse.
The pulse-weighted charge state averaged over time is about +12, demonstrating the severe electronic damage incurred during the x-ray pulse.

\begin{figure}
\includegraphics[scale=\figurescale]{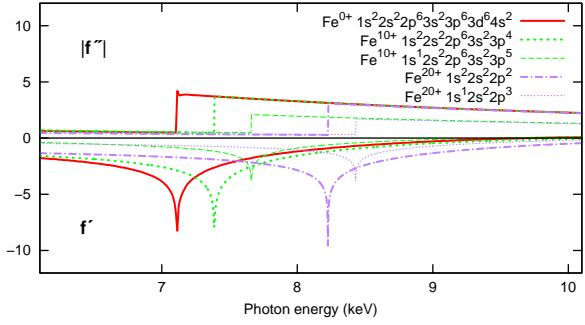}
\caption{(Color online) Dispersion corrections of atomic form factors for selected configurations of several charge states of Fe.}
\label{fig:form_factor}
\end{figure}

This electronic damage affects not only the coherent scattering atomic form factor but also its dispersion correction.
Near an inner-shell absorption edge, resonant elastic scattering causes the atomic form factor to depend on the photon energy $\omega$,
\begin{equation}
f(\mathbf{Q},\omega) = f^0(\mathbf{Q}) + f'(\omega) + i f''(\omega),
\end{equation}
where $\mathbf{Q}$ is the photon momentum transfer.
The \textsc{xatom} toolkit has been extended to compute the dispersion correction, $f' + i f''$~\cite{xatom}.
In Fig.~\ref{fig:form_factor}, one can see remarkable changes of the dispersion correction for different configurations and different charge states of Fe.
Both $f'$ and $f''$ have a singular position at the $K$-shell edge, which is shifted to a higher $\omega$ by $\sim$1~keV as the charge state increases.
The plotted curves in Fig.~\ref{fig:form_factor} correspond to the configurations of the ground state and the single-core-hole state (except for the neutral Fe) for given charge states.
Since the MAD phasing method is based on the dispersion correction of heavy elements, it is inevitably required to take into account the electronic damage dynamics and accompanying changes of the dispersion correction under intense x-ray pulses.

In the MAD phasing method, the Karle--Hendrickson equation~\cite{Karle80,Hendrickson85} represents a set of equations of scattering cross sections at several different wavelengths (photon energies).
The molecular scattering form factor is separated into normal and anomalous scattering terms and the phase information can be derived from their interferences.
In this Letter, we show that a Karle--Hendrickson-type equation exists in the high-intensity regime with extensive electronic damage on anomalous scatterers.

Let $P$ be any protein (or any macromolecule) whose structure we want to solve by coherent x-ray scattering.  
Let $H$ indicate heavy atoms and $N_H$ be the number of heavy atoms per macromolecule to be considered.
Note that $P$ excludes $H$.
Our assumption is that only heavy atoms scatter anomalously and undergo damage dynamics during an x-ray pulse.
It is justified by the fact that the photon energy of interest is near the inner-shell ionization threshold of heavy atoms and the photoabsorption cross section $\sigma$ of the heavy atom is much higher than that of the light atom for a given range of $\omega$.
For example, $\sigma_\text{Fe} / \sigma_\text{C} \approx 300$ at 8~keV and there is almost no dispersion effect on carbon (C) near this photon energy (see Fig.~S1(b) in Ref.~\cite{Generalized_KH}).
The scattering intensity (per unit solid angle) is evaluated by~\cite{Generalized_KH},
\begin{align}\label{eq:general}
\frac{d I(\mathbf{Q},\omega)}{d \Omega} & = 
\mathcal{F} C(\Omega) \int_{-\infty}^{\infty} \!\!\! dt \, g(t) \sum_I P_I(t) 
\nonumber \\ & \quad \times
\left| 
F^0_P(\mathbf{Q}) + \sum_{j=1}^{N_H} f_{I_j}(\mathbf{Q},\omega) e^{i \mathbf{Q} \cdot \mathbf{R}_j }
\right|^2,
\end{align}
where $j$ denotes a heavy atom index and $I$ indicates a global configuration index.
The global configuration for $N_H$ heavy atoms is given by $I=(I_1, I_2, \cdots, I_{N_H})$.
Here $I_j$ indicates the electronic configuration of the $j$-th heavy atom, which is located at position $\mathbf{R}_j$.
This electronic configuration provides, among other things, information on the charge state of the atom.
$P_I(t)$ is the population of the $I$-th configuration at time $t$.
It is assumed that the heavy atoms are ionized independently, so the population of $I$ is given by a product of individual populations, $P_I(t) = \Pi_{j=1}^{N_H} P_{I_j}(t)$.
$\mathcal{F}$ is the x-ray fluence and $g(t)$ is the normalized pulse envelope.
Then the x-ray flux is given by $\mathcal{F} g(t)$, which is assumed to be spatially uniform throughout the sample.
$C(\Omega)$ is a coefficient given by the polarization of the x-ray pulse.

In Eq.~(\ref{eq:general}), $F^0_P(\mathbf{Q})$ is the molecular form factor for the protein (without any dispersion correction) and our purpose is to solve its amplitude and phase, $F^0_P(\mathbf{Q}) = |F^0_P(\mathbf{Q})| \exp[{i \phi^0_P(\mathbf{Q})}]$.
$f_{I_j}(\mathbf{Q},\omega)$ is the atomic form factor (with the dispersion correction) of the $j$-th heavy atom in its $I_j$-th configuration.
It is most instructive to consider only one heavy atomic species.
We introduce a molecular form factor for undamaged heavy atoms,
\begin{equation}\label{eq:F^0_H}
F^0_H(\mathbf{Q}) = |F^0_H(\mathbf{Q})| e^{i \phi^0_H(\mathbf{Q})} = f^0_H(\mathbf{Q}) \sum_{j=1}^{N_H} e^{i \mathbf{Q} \cdot \mathbf{R}_j},
\end{equation}
where $f^0_H(\mathbf{Q})$ indicates the normal scattering atomic form factor for the ground-state configuration of the neutral heavy atom.

Now Eq.~(\ref{eq:general}) can be expanded to demonstrate the existence of a generalized Karle--Hendrickson equation~\cite{Generalized_KH},
\begin{align}\label{eq:generalized_KH}
\frac{d I(\mathbf{Q},\omega)}{d \Omega}
&
= \mathcal{F} C(\Omega) \Big[
\left| F^0_P(\mathbf{Q}) \right|^2
 + 
\left| F^0_H(\mathbf{Q}) \right|^2
\tilde{a}(\mathbf{Q},\omega)
\nonumber
\\
& + 
\left| F^0_P(\mathbf{Q}) \right|
\left| F^0_H(\mathbf{Q}) \right|
b(\mathbf{Q},\omega)
\cos\left( \phi^0_P(\mathbf{Q}) - \phi^0_H(\mathbf{Q}) \right) 
\nonumber
\\
& + 
\left| F^0_P(\mathbf{Q}) \right| 
\left| F^0_H(\mathbf{Q}) \right| 
c(\mathbf{Q},\omega)
\sin\left( \phi^0_P(\mathbf{Q}) - \phi^0_H(\mathbf{Q}) \right) 
\nonumber
\\
& + 
N_H 
\left| f^0_H(\mathbf{Q}) \right|^2
\left\lbrace a(\mathbf{Q},\omega) - \tilde{a}(\mathbf{Q},\omega) \right\rbrace
\Big],
\end{align}
where the new MAD coefficients depending on $\mathbf{Q}$ and $\omega$ are defined by
\begin{subequations}\label{eq:coefficient}
\begin{align}
\label{eq:a}
a(\mathbf{Q},\omega) &= \frac{1}{\left\lbrace f^0_H(\mathbf{Q}) \right\rbrace^2} \sum_{I_H} \bar{P}_{I_H} \left| f_{I_H}(\mathbf{Q},\omega) \right|^2, 
\\
\label{eq:b}
b(\mathbf{Q},\omega) &= \frac{2}{f^0_H(\mathbf{Q})} \sum_{I_H} \bar{P}_{I_H} \left\lbrace f^0_{I_H}(\mathbf{Q}) + f'_{I_H}(\omega) \right\rbrace,
\\
\label{eq:c}
c(\mathbf{Q},\omega) &= \frac{2}{f^0_H(\mathbf{Q})} \sum_{I_H} \bar{P}_{I_H} f''_{I_H}(\omega),
\\
\label{eq:tilde_a}
\tilde{a}(\mathbf{Q},\omega) &= \frac{1}{\left\lbrace f^0_H(\mathbf{Q}) \right\rbrace^2} \int_{-\infty}^{\infty} \!\! dt \, g(t) \left| \tilde{f}_{H}(\mathbf{Q}, \omega, t) \right|^2.
\end{align}
\end{subequations}
Here $I_H$ indicates the electronic configuration of the heavy atom species and $\bar{P}_{I_H} = \int_{-\infty}^{\infty} dt \, g(t) P_{I_H}(t)$ is the pulse-weighted averaged population for the $I_H$-th configuration.
The new MAD coefficients from Eq.~(\ref{eq:a}) to Eq.~(\ref{eq:tilde_a}) are atom-specific and must be calculated with electronic damage dynamics and configuration-specific atomic form factors.
The coefficient $a$ is an incoherent average of $|f_{I_H}|^2$ with $\bar{P}_{I_H}$.
The coefficients $b$ and $c$ are the real and imaginary components of the averaged atomic form factor, respectively.
The coefficient $\tilde{a}$ in Eq.~(\ref{eq:tilde_a}) is obtained through a \textit{dynamical} form factor defined by
\begin{equation}
\tilde{f}_{H}(\mathbf{Q}, \omega, t) = \sum_{I_H} P_{I_H}(t) f_{I_H}(\mathbf{Q},\omega),
\end{equation}
which is a coherent average of the configuration-specific form factors over $I_H$ at a given time $t$.
This $\tilde{a}$ coefficient thus represents the effective scattering strength of the heavy atom.
In contrast to the original Karle--Hendrickson equation, Eq.~(\ref{eq:generalized_KH}) is separated into light atoms ($P$) and heavy atoms ($H$) because both electronic damage and anomalous scattering are treated exclusively on $H$.
If only the ground-state configuration is considered, i.e., no electronic damage occurs, then $a = \tilde{a}$ and Eqs.~(\ref{eq:generalized_KH}) and (\ref{eq:coefficient}) are reduced to the original Karle--Hendrickson equation except for the separation of $P$ and $H$.

This generalized Karle--Hendrickson equation constitutes a set of equations with different $\omega$ at every $\mathbf{Q}$.
In Eq.~(\ref{eq:generalized_KH}) there are three unknowns: $\left| F^0_P(\mathbf{Q}) \right|$, $\left| F^0_H(\mathbf{Q}) \right|$, and $\phi^0_P(\mathbf{Q}) - \phi^0_H(\mathbf{Q})$ for a given $\mathbf{Q}$.
With three or more different $\omega$, those unknowns can be solved by the least-square method~\cite{Karle89,Hendrickson88}.
Combined with Patterson or direct methods~\cite{Hauptman86,Karle86}, the amplitude and phase of heavy atoms can be determined, so two unknowns of $\left| F^0_P(\mathbf{Q}) \right|$ and $\phi^0_P(\mathbf{Q})$ are to be solved with two different $\omega$.
Once all amplitudes and phases of $P$ and $H$ are determined, it is straightforward to construct the total structure of $T=P+H$.
To obtain non-trivial solutions from the least-square method, the contrast between the coefficients at two different $\omega$ must be non-zero.
This condition is fulfilled even in the presence of severe electronic damage as shown in the following discussion.
We emphasize that, if the MAD coefficients are predetermined experimentally or theoretically, then one can solve the structure (amplitude and phase) from diffraction measurements directly, without any iterative phase retrieval algorithms~\cite{Millane90}.

\begin{figure}
\centering
\includegraphics[scale=\figurescale]{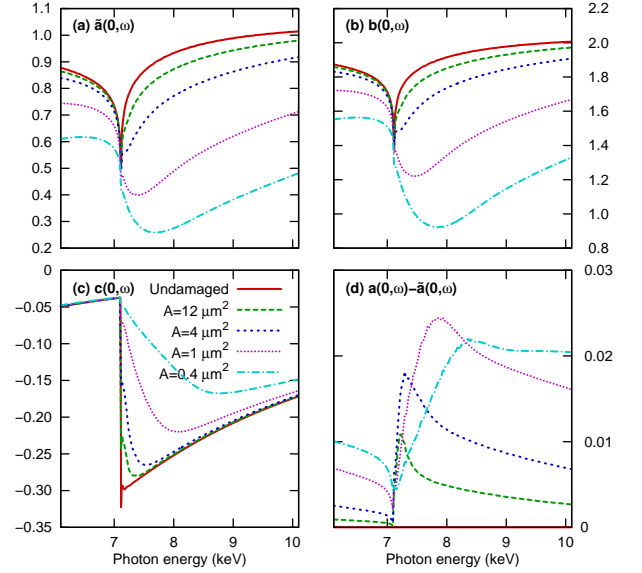}
\caption{(Color online) Coefficients in the generalized Karle--Hendrickson equation for Fe as a function of the photon energy.  The fluence is given by $2$$\times$$10^{12}~\text{photons} / A$ where $A$ is the focal spot area.}
\label{fig:abc}
\end{figure}

Let us consider Fe atoms embedded in a protein and then radiate an x-ray pulse of 2$\times$10$^{12}$ photons and 10~fs FWHM into the sample.
Figure~\ref{fig:abc} displays $\tilde{a}$, $b$, $c$, and $(a-\tilde{a})$ for the forward direction ($\mathbf{Q}=\mathbf{0}$) computed by the extended \textsc{xatom} toolkit~\cite{xatom}.
The fluence $\mathcal{F}$ is given by $2$$\times$$10^{12}~\text{photons} / A$ where $A$ is the focal spot area.
When high charge states are generated by ionization dynamics, the scattering strength is lowered due to the reduced number of scattering electrons and the change of the dispersion correction.
The degree of lowering in $\tilde{a}$ and $b$ shows different behaviors below and above the neutral Fe edge. 
Below the edge, the scattering strength is less lowered than above the edge because ionization dynamics are dominantly initiated by $L$-shell ionization whose cross section is 8 times smaller than that of $K$-shell ionization~\cite{Bethe08}.
Above the edge, $K$-shell ionization channels are open and lead to further cascade decays, stripping off more electrons.
As a result, $\tilde{a}$ and $b$ are dramatically bleached out and their minimum is deepened and broadened.
The absolute value of $c$, which corresponds to the averaged absorption cross section, is decreased as the fluence increases.

This bleaching effect on the scattering strength is beneficial to the phasing problem in two ways.
First, the contrast of the coefficients to be exploited in the MAD method is enhanced.
Even though the scattering strength is lowered for all $\omega$, Figs.~\ref{fig:abc}(a)--(c) clearly show that the low intensity cases (long-dashed and short-dashed lines) display a contrast similar to the conventional MAD method (solid lines).
For the high intensity cases (dotted and dash-dotted lines), the contrast in $\tilde{a}$ and $b$ becomes even larger when $\omega$ is chosen below the edge and around the minimum.
The contrast in $c$ is reduced to some extent but is not completely eliminated.
It is worthwhile to note that broadening of the edge at high intensity makes precision of $\omega$ less important in experiments.
Second, it brings an alternative phasing method similar to single isomorphic replacement (SIR)~\cite{Taylor03} or radiation-damage induced phasing (RIP)~\cite{Ravelli03}.
By choosing one $\omega$ below and one $\omega$ above the edge, one can create two datasets that differ only in the scattering strength of the heavy atoms, and then solve the rest of the structure by density modification.
In this method, there is neither atomic replacement in sample preparation like SIR nor chemical rearrangement during the x-ray pulses like RIP.
Therefore, the rest of the structure remains invariant in the two different datasets.

Now we discuss experimental implementation of the generalized version of the MAD phasing method.
We used a Gaussian pulse in the above calculations. 
However, when the scattering strength of $dI / d\Omega$ is measured at a particular $\mathbf{Q}$ and $\omega$, the pulse shape $g(t)$ and the fluence $\mathcal{F}$ at a given position $\mathbf{x}$ in the x-ray beam may differ from shot to shot.
Following the procedure in Ref.~\cite{Rohringer07}, we numerically confirmed that $\langle dI[ \mathcal{F}(\mathbf{x}) g(t) ] / d\Omega \rangle \approx dI[ \langle \mathcal{F}(\mathbf{x}) g(t) \rangle ] / d\Omega $ to within 3\%, where $\langle~\cdot~\rangle$ denotes an ensemble average.
Then the total signal can be obtained by integrating over the interaction volume, $\int d^3 x \, dI \left[ \langle \mathcal{F}(\mathbf{x}) g(t) \rangle \right] / d\Omega \times n_\text{mol}(\mathbf{x})$, where $n_\text{mol}(\mathbf{x})$ is the molecular number density.
In this process, the MAD coefficients are calculated with given $\langle \mathcal{F}g(t) \rangle$, and the basic structure of Eq.~(\ref{eq:generalized_KH}) remains unchanged.

In our model, resonant absorption processes and shake-up/shake-off processes~\cite{Persson01} are neglected.
They tend to generate further high charge states, so the contrast enhancement and the bleaching effect would remain after inclusion of these processes.
We note that the effect of impact ionization~\cite{Ziaja05} on coherent diffractive imaging may be suppressed by using a sufficiently short x-ray pulse~\cite{Son11a}.

The structure of the generalized Karle--Hendrickson equation [Eq.~(\ref{eq:generalized_KH})] can be fully functional for phasing of nanocrystals, which are of current interest for structural determination with x-ray FELs~\cite{Mancuso09,Chapman11}.
In Eq.~(\ref{eq:F^0_H}), $F^0_H$ contains the structure factors of the heavy atoms.
In the case of crystals, the heavy atoms are regularly located and can contribute to the Bragg peaks when satisfying $\mathbf{Q} \cdot \left( \mathbf{R}_i - \mathbf{R}_j \right) = 2\pi n$ ($n$: integer) for all $i$ and $j$.
In Eq.~(\ref{eq:tilde_a}), $\tilde{a}$ is expressed with the coherent average over configurations, and $| F^0_H |^2 $ from Eq.~(\ref{eq:F^0_H}) is expressed with the coherent summation over heavy atoms.
Therefore, $ | F^0_H |^2 \tilde{a} = \int_{-\infty}^{\infty} dt \, g(t) \left| \tilde{f}_{H}(\mathbf{Q}, \omega, t) \sum_{j=1}^{N_H} \exp[ i \mathbf{Q} \cdot \mathbf{R}_j ] \right|^2 $ implies that all heavy atoms are described by the same \textit{dynamical} form factor.
This term is then responsible for the Bragg peaks ($\propto N_H^2$).
On the other hand, the term $N_H | f^0_H |^2 ( a - \tilde{a})$ represents fluctuations from all different configurations induced by electronic damage dynamics, corresponding to the diffuse background ($\propto N_H$).
As shown in Fig.~\ref{fig:abc}(d), $( a - \tilde{a})$ increases as the fluence increases.
However, it is an order of magnitude smaller than $\tilde{a}$ and not confined to the Bragg peaks, implying that the high x-ray intensity does not fully destroy the coherent signals.

In conclusion, 
we have proposed the MAD phasing method in extreme conditions of ionizing x-ray radiations.
We assume that the scattering factors of the light atoms of the protein do not vary significantly over the measured range of x-ray frequencies, and that these atoms have normal scattering and no ionization.
It is also assumed that the heavy atoms are ionized independently and only one type of heavy atoms is considered.
We believe that the method should work even if these assumptions are removed, because the most important consequence of high-intensity x-ray irradiation --- multiple ionization of the heavy atomic species --- has been fully taken into account.
We have combined electronic response at the atomic level and molecular imaging during intense x-ray pulses, and demonstrated the existence of a generalized Karle--Hendrickson equation for the MAD method at high x-ray intensity.
The relevant coefficients to be used in the MAD method have been formulated and calculated with damage dynamics and accompanying changes of the dispersion correction.
We have shown that the generalized equation is still applicable to the phase problem even in the presence of severe radiation damage.
The bleaching effect on the scattering strength of heavy atoms, which unexpectedly enhances the coefficient contrast in the MAD method, can be beneficial to phasing.
Our study opens up a new opportunity of solving the phase problem in femtosecond nanocrystallography with x-ray FELs.


\begin{acknowledgments}
The authors thank Stefan Pabst and Dr.\ Huijong Han for helpful discussions and Urs B\"ultemeier for proof reading.
\end{acknowledgments}




\begin{thebibliography}{35}%
\makeatletter
\providecommand \@ifxundefined [1]{%
 \@ifx{#1\undefined}
}%
\providecommand \@ifnum [1]{%
 \ifnum #1\expandafter \@firstoftwo
 \else \expandafter \@secondoftwo
 \fi
}%
\providecommand \@ifx [1]{%
 \ifx #1\expandafter \@firstoftwo
 \else \expandafter \@secondoftwo
 \fi
}%
\providecommand \natexlab [1]{#1}%
\providecommand \enquote  [1]{``#1''}%
\providecommand \bibnamefont  [1]{#1}%
\providecommand \bibfnamefont [1]{#1}%
\providecommand \citenamefont [1]{#1}%
\providecommand \href@noop [0]{\@secondoftwo}%
\providecommand \href [0]{\begingroup \@sanitize@url \@href}%
\providecommand \@href[1]{\@@startlink{#1}\@@href}%
\providecommand \@@href[1]{\endgroup#1\@@endlink}%
\providecommand \@sanitize@url [0]{\catcode `\\12\catcode `\$12\catcode
  `\&12\catcode `\#12\catcode `\^12\catcode `\_12\catcode `\%12\relax}%
\providecommand \@@startlink[1]{}%
\providecommand \@@endlink[0]{}%
\providecommand \url  [0]{\begingroup\@sanitize@url \@url }%
\providecommand \@url [1]{\endgroup\@href {#1}{\urlprefix }}%
\providecommand \urlprefix  [0]{URL }%
\providecommand \Eprint [0]{\href }%
\providecommand \doibase [0]{http://dx.doi.org/}%
\providecommand \selectlanguage [0]{\@gobble}%
\providecommand \bibinfo  [0]{\@secondoftwo}%
\providecommand \bibfield  [0]{\@secondoftwo}%
\providecommand \translation [1]{[#1]}%
\providecommand \BibitemOpen [0]{}%
\providecommand \bibitemStop [0]{}%
\providecommand \bibitemNoStop [0]{.\EOS\space}%
\providecommand \EOS [0]{\spacefactor3000\relax}%
\providecommand \BibitemShut  [1]{\csname bibitem#1\endcsname}%
\let\auto@bib@innerbib\@empty
\bibitem [{\citenamefont {Dauter}(2006)}]{Dauter06}%
  \BibitemOpen
  \bibfield  {author} {\bibinfo {author} {\bibfnamefont {Z.}~\bibnamefont
  {Dauter}},\ }\href@noop {} {\bibfield  {journal} {\bibinfo  {journal} {Acta
  Cryst.}\ }\textbf {\bibinfo {volume} {D62}},\ \bibinfo {pages} {1} (\bibinfo
  {year} {2006})}\BibitemShut {NoStop}%
\bibitem [{\citenamefont {Karle}\ and\ \citenamefont
  {Hauptman}(1950)}]{Karle50}%
  \BibitemOpen
  \bibfield  {author} {\bibinfo {author} {\bibfnamefont {J.}~\bibnamefont
  {Karle}}\ and\ \bibinfo {author} {\bibfnamefont {H.}~\bibnamefont
  {Hauptman}},\ }\href@noop {} {\bibfield  {journal} {\bibinfo  {journal} {Acta
  Cryst.}\ }\textbf {\bibinfo {volume} {3}},\ \bibinfo {pages} {181} (\bibinfo
  {year} {1950})}\BibitemShut {NoStop}%
\bibitem [{\citenamefont {Taylor}(2003)}]{Taylor03}%
  \BibitemOpen
  \bibfield  {author} {\bibinfo {author} {\bibfnamefont {G.}~\bibnamefont
  {Taylor}},\ }\href@noop {} {\bibfield  {journal} {\bibinfo  {journal} {Acta
  Cryst.}\ }\textbf {\bibinfo {volume} {D59}},\ \bibinfo {pages} {1881}
  (\bibinfo {year} {2003})}\BibitemShut {NoStop}%
\bibitem [{\citenamefont {Guss}\ \emph {et~al.}(1988)\citenamefont {Guss},
  \citenamefont {Merritt}, \citenamefont {Phizackerley}, \citenamefont
  {Hedman}, \citenamefont {Murata}, \citenamefont {Hodgson},\ and\
  \citenamefont {Freeman}}]{Guss88}%
  \BibitemOpen
  \bibfield  {author} {\bibinfo {author} {\bibfnamefont {J.~M.}\ \bibnamefont
  {Guss}} \emph{et~al.},\ }\href@noop {} {\bibfield
  {journal} {\bibinfo  {journal} {Science}\ }\textbf {\bibinfo {volume}
  {241}},\ \bibinfo {pages} {806} (\bibinfo {year} {1988})}\BibitemShut
  {NoStop}%
\bibitem [{\citenamefont {Hendrickson}\ \emph {et~al.}(1989)\citenamefont
  {Hendrickson}, \citenamefont {P{\"a}hler}, \citenamefont {Smith},
  \citenamefont {Satow}, \citenamefont {Merritt},\ and\ \citenamefont
  {Phizackerley}}]{Hendrickson89}%
  \BibitemOpen
  \bibfield  {author} {\bibinfo {author} {\bibfnamefont {W.~A.}\ \bibnamefont
  {Hendrickson}} \emph{et~al.},\
  }\href@noop {} {\bibfield  {journal} {\bibinfo  {journal} {Proc. Natl. Acad.
  Sci. U. S. A.}\ }\textbf {\bibinfo {volume} {86}},\ \bibinfo {pages} {2190}
  (\bibinfo {year} {1989})}\BibitemShut {NoStop}%
\bibitem [{\citenamefont {Hendrickson}(1991)}]{Hendrickson91}%
  \BibitemOpen
  \bibfield  {author} {\bibinfo {author} {\bibfnamefont {W.~A.}\ \bibnamefont
  {Hendrickson}},\ }\href@noop {} {\bibfield  {journal} {\bibinfo  {journal}
  {Science}\ }\textbf {\bibinfo {volume} {254}},\ \bibinfo {pages} {51}
  (\bibinfo {year} {1991})}\BibitemShut {NoStop}%
\bibitem [{\citenamefont {McNeil}\ and\ \citenamefont
  {Thompson}(2010)}]{McNeil10}%
  \BibitemOpen
  \bibfield  {author} {\bibinfo {author} {\bibfnamefont {B.~W.~J.}\
  \bibnamefont {McNeil}}\ and\ \bibinfo {author} {\bibfnamefont {N.~R.}\
  \bibnamefont {Thompson}},\ }\href@noop {} {\bibfield  {journal} {\bibinfo
  {journal} {Nature Photon.}\ }\textbf {\bibinfo {volume} {4}},\ \bibinfo
  {pages} {814} (\bibinfo {year} {2010})}\BibitemShut {NoStop}%
\bibitem [{\citenamefont {Gaffney}\ and\ \citenamefont
  {Chapman}(2007)}]{Gaffney07}%
  \BibitemOpen
  \bibfield  {author} {\bibinfo {author} {\bibfnamefont {K.~J.}\ \bibnamefont
  {Gaffney}}\ and\ \bibinfo {author} {\bibfnamefont {H.~N.}\ \bibnamefont
  {Chapman}},\ }\href@noop {} {\bibfield  {journal} {\bibinfo  {journal}
  {Science}\ }\textbf {\bibinfo {volume} {316}},\ \bibinfo {pages} {1444}
  (\bibinfo {year} {2007})}\BibitemShut {NoStop}%
\bibitem [{\citenamefont {Neutze}\ \emph {et~al.}(2000)\citenamefont {Neutze},
  \citenamefont {Wouts}, \citenamefont {{van der Spoel}}, \citenamefont
  {Weckert},\ and\ \citenamefont {Hajdu}}]{Neutze00}%
  \BibitemOpen
  \bibfield  {author} {\bibinfo {author} {\bibfnamefont {R.}~\bibnamefont
  {Neutze}} \emph {et~al.},\ }\href@noop {}
  {\bibfield  {journal} {\bibinfo  {journal} {Nature}\ }\textbf {\bibinfo
  {volume} {406}},\ \bibinfo {pages} {752} (\bibinfo {year}
  {2000})}\BibitemShut {NoStop}%
\bibitem [{\citenamefont {Chapman}\ \emph {et~al.}(2006)\citenamefont
  {Chapman}, \citenamefont {Barty}, \citenamefont {Bogan}, \citenamefont
  {Boutet}, \citenamefont {Frank}, \citenamefont {Hau-Riege}, \citenamefont
  {Marchesini}, \citenamefont {Woods}, \citenamefont {Bajt}, \citenamefont
  {Benner}, \citenamefont {London}, \citenamefont {Pl{\"o}njes}, \citenamefont
  {Kuhlmann}, \citenamefont {Treusch}, \citenamefont {D{\"u}sterer},
  \citenamefont {Tschentscher}, \citenamefont {Schneider}, \citenamefont
  {Spiller}, \citenamefont {M{\"o}ller}, \citenamefont {Bostedt}, \citenamefont
  {Hoener}, \citenamefont {Shapiro}, \citenamefont {Hodgson}, \citenamefont
  {Van~der Spoel}, \citenamefont {Burmeister}, \citenamefont {Bergh},
  \citenamefont {Caleman}, \citenamefont {Huldt}, \citenamefont {Seibert},
  \citenamefont {Maia}, \citenamefont {Lee}, \citenamefont {Sz{\"o}ke},
  \citenamefont {Timneanu},\ and\ \citenamefont {Hajdu}}]{Chapman06}%
  \BibitemOpen
  \bibfield  {author} {\bibinfo {author} {\bibfnamefont {H.~N.}\ \bibnamefont
  {Chapman}} \emph {et~al.},\ }\href@noop {}
  {\bibfield  {journal} {\bibinfo  {journal} {Nature Phys.}\ }\textbf {\bibinfo
  {volume} {2}},\ \bibinfo {pages} {839} (\bibinfo {year} {2006})}\BibitemShut
  {NoStop}%
\bibitem [{\citenamefont {Chapman}\ and\ \citenamefont
  {Nugent}(2010)}]{Chapman10}%
  \BibitemOpen
  \bibfield  {author} {\bibinfo {author} {\bibfnamefont {H.~N.}\ \bibnamefont
  {Chapman}}\ and\ \bibinfo {author} {\bibfnamefont {K.~A.}\ \bibnamefont
  {Nugent}},\ }\href@noop {} {\bibfield  {journal} {\bibinfo  {journal} {Nature
  Photon.}\ }\textbf {\bibinfo {volume} {4}},\ \bibinfo {pages} {833} (\bibinfo
  {year} {2010})}\BibitemShut {NoStop}%
\bibitem [{\citenamefont {Mancuso}\ \emph {et~al.}(2009)\citenamefont
  {Mancuso}, \citenamefont {Schropp}, \citenamefont {Reime}, \citenamefont
  {Stadler}, \citenamefont {Singer}, \citenamefont {Gulden}, \citenamefont
  {Streit-Nierobisch}, \citenamefont {Gutt}, \citenamefont {Gr\"ubel},
  \citenamefont {Feldhaus}, \citenamefont {Staier}, \citenamefont {Barth},
  \citenamefont {Rosenhahn}, \citenamefont {Grunze}, \citenamefont {Nisius},
  \citenamefont {Wilhein}, \citenamefont {Stickler}, \citenamefont {Stillrich},
  \citenamefont {Fr\"omter}, \citenamefont {Oepen}, \citenamefont {Martins},
  \citenamefont {Pfau}, \citenamefont {G\"unther}, \citenamefont {K\"onnecke},
  \citenamefont {Eisebitt}, \citenamefont {Faatz}, \citenamefont
  {Guerassimova}, \citenamefont {Honkavaara}, \citenamefont {Kocharyan},
  \citenamefont {Treusch}, \citenamefont {Saldin}, \citenamefont {Schreiber},
  \citenamefont {Schneidmiller}, \citenamefont {Yurkov}, \citenamefont
  {Weckert},\ and\ \citenamefont {Vartanyants}}]{Mancuso09}%
  \BibitemOpen
  \bibfield  {author} {\bibinfo {author} {\bibfnamefont {A.~P.}\ \bibnamefont
  {Mancuso}} \emph {et~al.},\ }\href@noop {} {\bibfield  {journal} {\bibinfo  {journal}
  {Phys. Rev. Lett.}\ }\textbf {\bibinfo {volume} {102}},\ \bibinfo {pages}
  {035502} (\bibinfo {year} {2009})}\BibitemShut {NoStop}%
\bibitem [{\citenamefont {Chapman}\ \emph {et~al.}(2011)\citenamefont
  {Chapman}, \citenamefont {Fromme}, \citenamefont {Barty}, \citenamefont
  {White}, \citenamefont {Kirian}, \citenamefont {Aquila}, \citenamefont
  {Hunter}, \citenamefont {Schulz}, \citenamefont {DePonte}, \citenamefont
  {Weierstall}, \citenamefont {Doak}, \citenamefont {Maia}, \citenamefont
  {Martin}, \citenamefont {Schlichting}, \citenamefont {Lomb}, \citenamefont
  {Coppola}, \citenamefont {Shoeman}, \citenamefont {Epp}, \citenamefont
  {Hartmann}, \citenamefont {Rolles}, \citenamefont {Rudenko}, \citenamefont
  {Foucar}, \citenamefont {Kimmel}, \citenamefont {Weidenspointner},
  \citenamefont {Holl}, \citenamefont {Liang}, \citenamefont {Barthelmess},
  \citenamefont {Caleman}, \citenamefont {Boutet}, \citenamefont {Bogan},
  \citenamefont {Krzywinski}, \citenamefont {Bostedt}, \citenamefont {Bajt},
  \citenamefont {Gumprecht}, \citenamefont {Rudek}, \citenamefont {Erk},
  \citenamefont {Schmidt}, \citenamefont {H{\"o}mke}, \citenamefont {Reich},
  \citenamefont {Pietschner}, \citenamefont {Str{\"u}der}, \citenamefont
  {Hauser}, \citenamefont {Gorke}, \citenamefont {Ullrich}, \citenamefont
  {Herrmann}, \citenamefont {Schaller}, \citenamefont {Schopper}, \citenamefont
  {Soltau}, \citenamefont {K{\"u}hnel}, \citenamefont {Messerschmidt},
  \citenamefont {Bozek}, \citenamefont {Hau-Riege}, \citenamefont {Frank},
  \citenamefont {Hampton}, \citenamefont {Sierra}, \citenamefont {Starodub},
  \citenamefont {Williams}, \citenamefont {Hajdu}, \citenamefont {Timneanu},
  \citenamefont {Seibert}, \citenamefont {Andreasson}, \citenamefont {Rocker},
  \citenamefont {J{\"o}nsson}, \citenamefont {Svenda}, \citenamefont {Stern},
  \citenamefont {Nass}, \citenamefont {Andritschke}, \citenamefont
  {Schr{\"o}ter}, \citenamefont {Krasniqi}, \citenamefont {Bott}, \citenamefont
  {Schmidt}, \citenamefont {Wang}, \citenamefont {Grotjohann}, \citenamefont
  {Holton}, \citenamefont {Barends}, \citenamefont {Neutze}, \citenamefont
  {Marchesini}, \citenamefont {Fromme}, \citenamefont {Schorb}, \citenamefont
  {Rupp}, \citenamefont {Adolph}, \citenamefont {Gorkhover}, \citenamefont
  {Andersson}, \citenamefont {Hirsemann}, \citenamefont {Potdevin},
  \citenamefont {Graafsma}, \citenamefont {Nilsson},\ and\ \citenamefont
  {Spence}}]{Chapman11}%
  \BibitemOpen
  \bibfield  {author} {\bibinfo {author} {\bibfnamefont {H.~N.}\ \bibnamefont
  {Chapman}} \emph {et~al.},\ }\href@noop {} {\bibfield
   {journal} {\bibinfo  {journal} {Nature}\ }\textbf {\bibinfo {volume}
  {470}},\ \bibinfo {pages} {73} (\bibinfo {year} {2011})}\BibitemShut
  {NoStop}%
\bibitem [{\citenamefont {Henderson}(1995)}]{Henderson95}%
  \BibitemOpen
  \bibfield  {author} {\bibinfo {author} {\bibfnamefont {R.}~\bibnamefont
  {Henderson}},\ }\href@noop {} {\bibfield  {journal} {\bibinfo  {journal} {Q.
  Rev. Biophys.}\ }\textbf {\bibinfo {volume} {28}},\ \bibinfo {pages} {171}
  (\bibinfo {year} {1995})}\BibitemShut {NoStop}%
\bibitem [{\citenamefont {Howells}\ \emph {et~al.}(2009)\citenamefont
  {Howells}, \citenamefont {Beetz}, \citenamefont {Chapman}, \citenamefont
  {Cui}, \citenamefont {Holton}, \citenamefont {Jacobsen}, \citenamefont
  {Kirz}, \citenamefont {Lima}, \citenamefont {Marchesini}, \citenamefont
  {Miao}, \citenamefont {Sayre}, \citenamefont {Shapiro}, \citenamefont
  {Spence},\ and\ \citenamefont {Starodub}}]{Howells09}%
  \BibitemOpen
  \bibfield  {author} {\bibinfo {author} {\bibfnamefont {M.~R.}\ \bibnamefont
  {Howells}} \emph {et~al.},\ }\href@noop {} {\bibfield
  {journal} {\bibinfo  {journal} {J. Electron Spectrosc. Relat. Phenom.}\
  }\textbf {\bibinfo {volume} {170}},\ \bibinfo {pages} {4} (\bibinfo {year}
  {2009})}\BibitemShut {NoStop}%
\bibitem [{\citenamefont {Hau-Riege}\ \emph {et~al.}(2004)\citenamefont
  {Hau-Riege}, \citenamefont {London},\ and\ \citenamefont
  {Szoke}}]{Hau-Riege04}%
  \BibitemOpen
  \bibfield  {author} {\bibinfo {author} {\bibfnamefont {S.~P.}\ \bibnamefont
  {Hau-Riege}}, \bibinfo {author} {\bibfnamefont {R.~A.}\ \bibnamefont
  {London}},\ and\ \bibinfo {author} {\bibfnamefont {A.}~\bibnamefont
  {Szoke}},\ }\href@noop {} {\bibfield  {journal} {\bibinfo  {journal} {Phys.
  Rev. E}\ }\textbf {\bibinfo {volume} {69}},\ \bibinfo {pages} {051906}
  (\bibinfo {year} {2004})}\BibitemShut {NoStop}%
\bibitem [{\citenamefont {Quiney}\ and\ \citenamefont
  {Nugent}(2011)}]{Quiney11}%
  \BibitemOpen
  \bibfield  {author} {\bibinfo {author} {\bibfnamefont {H.~M.}\ \bibnamefont
  {Quiney}}\ and\ \bibinfo {author} {\bibfnamefont {K.~A.}\ \bibnamefont
  {Nugent}},\ }\href@noop {} {\bibfield  {journal} {\bibinfo  {journal} {Nature
  Phys.}\ }\textbf {\bibinfo {volume} {7}},\ \bibinfo {pages} {142} (\bibinfo
  {year} {2011})}\BibitemShut {NoStop}%
\bibitem [{\citenamefont {Son}\ \emph {et~al.}(2011)\citenamefont {Son},
  \citenamefont {Young},\ and\ \citenamefont {Santra}}]{Son11a}%
  \BibitemOpen
  \bibfield  {author} {\bibinfo {author} {\bibfnamefont {S.-K.}\ \bibnamefont
  {Son}}, \bibinfo {author} {\bibfnamefont {L.}~\bibnamefont {Young}},\ and\
  \bibinfo {author} {\bibfnamefont {R.}~\bibnamefont {Santra}},\ }\href@noop {}
  {\bibfield  {journal} {\bibinfo  {journal} {Phys. Rev. A}\ }\textbf {\bibinfo
  {volume} {83}},\ \bibinfo {pages} {033402} (\bibinfo {year}
  {2011})}\BibitemShut {NoStop}%
\bibitem [{\citenamefont {Ravelli}\ \emph {et~al.}(2005)\citenamefont
  {Ravelli}, \citenamefont {Nanao}, \citenamefont {Lovering}, \citenamefont
  {White},\ and\ \citenamefont {McSweeney}}]{Ravelli05}%
  \BibitemOpen
  \bibfield  {author} {\bibinfo {author} {\bibfnamefont {R.~B.~G.}\
  \bibnamefont {Ravelli}} \emph {et~al.},\
  }\href@noop {} {\bibfield  {journal} {\bibinfo  {journal} {J. Synchrotron
  Radiat.}\ }\textbf {\bibinfo {volume} {12}},\ \bibinfo {pages} {276}
  (\bibinfo {year} {2005})}\BibitemShut {NoStop}%
\bibitem [{\citenamefont {Young}\ \emph {et~al.}(2010)\citenamefont {Young},
  \citenamefont {Kanter}, \citenamefont {Kr{\"a}ssig}, \citenamefont {Li},
  \citenamefont {March}, \citenamefont {Pratt}, \citenamefont {Santra},
  \citenamefont {Southworth}, \citenamefont {Rohringer}, \citenamefont
  {DiMauro}, \citenamefont {Doumy}, \citenamefont {Roedig}, \citenamefont
  {Berrah}, \citenamefont {Fang}, \citenamefont {Hoener}, \citenamefont
  {Bucksbaum}, \citenamefont {Cryan}, \citenamefont {Ghimire}, \citenamefont
  {Glownia}, \citenamefont {Reis}, \citenamefont {Bozek}, \citenamefont
  {Bostedt},\ and\ \citenamefont {Messerschmidt}}]{Young10}%
  \BibitemOpen
  \bibfield  {author} {\bibinfo {author} {\bibfnamefont {L.}~\bibnamefont
  {Young}} \emph {et~al.},\ }\href@noop {}
  {\bibfield  {journal} {\bibinfo  {journal} {Nature}\ }\textbf {\bibinfo
  {volume} {466}},\ \bibinfo {pages} {56} (\bibinfo {year} {2010})}\BibitemShut
  {NoStop}%
\bibitem [{\citenamefont {Carlson}\ \emph {et~al.}(1966)\citenamefont
  {Carlson}, \citenamefont {Hunt},\ and\ \citenamefont {Krause}}]{Carlson66}%
  \BibitemOpen
  \bibfield  {author} {\bibinfo {author} {\bibfnamefont {T.~A.}\ \bibnamefont
  {Carlson}}, \bibinfo {author} {\bibfnamefont {W.~E.}\ \bibnamefont {Hunt}}, \
  and\ \bibinfo {author} {\bibfnamefont {M.~O.}\ \bibnamefont {Krause}},\
  }\href@noop {} {\bibfield  {journal} {\bibinfo  {journal} {Phys. Rev.}\
  }\textbf {\bibinfo {volume} {151}},\ \bibinfo {pages} {41} (\bibinfo {year}
  {1966})}\BibitemShut {NoStop}%
\bibitem{xatom}
  S.-K.\ Son and R.\ Santra, \textsc{xatom} --- an integrated toolkit for x-ray and atomic physics, CFEL, DESY, Hamburg, Germany, 2011, Rev.\ 398.
\bibitem [{\citenamefont {Karle}(1980)}]{Karle80}%
  \BibitemOpen
  \bibfield  {author} {\bibinfo {author} {\bibfnamefont {J.}~\bibnamefont
  {Karle}},\ }\href@noop {} {\bibfield  {journal} {\bibinfo  {journal} {Int. J.
  Quant. Chem. Quant. Bio. Symp.}\ }\textbf {\bibinfo {volume} {7}},\ \bibinfo
  {pages} {357} (\bibinfo {year} {1980})}\BibitemShut {NoStop}%
\bibitem [{\citenamefont {Hendrickson}(1985)}]{Hendrickson85}%
  \BibitemOpen
  \bibfield  {author} {\bibinfo {author} {\bibfnamefont {W.~A.}\ \bibnamefont
  {Hendrickson}},\ }\href@noop {} {\bibfield  {journal} {\bibinfo  {journal}
  {Trans. Am. Crystalgr. Assoc.}\ }\textbf {\bibinfo {volume} {21}},\ \bibinfo
  {pages} {11} (\bibinfo {year} {1985})}\BibitemShut {NoStop}%
\bibitem{Generalized_KH}%
  \BibitemOpen
  See Supplemental Material at [URL will be inserted by publisher] for a detailed derivation
  \BibitemShut {NoStop}%
\bibitem [{\citenamefont {Karle}(1989)}]{Karle89}%
  \BibitemOpen
  \bibfield  {author} {\bibinfo {author} {\bibfnamefont {J.}~\bibnamefont
  {Karle}},\ }\href@noop {} {\bibfield  {journal} {\bibinfo  {journal} {Acta
  Cryst.}\ }\textbf {\bibinfo {volume} {A45}},\ \bibinfo {pages} {303}
  (\bibinfo {year} {1989})}\BibitemShut {NoStop}%
\bibitem [{\citenamefont {Hendrickson}\ \emph {et~al.}(1988)\citenamefont
  {Hendrickson}, \citenamefont {Smith}, \citenamefont {Phizackerley},\ and\
  \citenamefont {Merritt}}]{Hendrickson88}%
  \BibitemOpen
  \bibfield  {author} {\bibinfo {author} {\bibfnamefont {W.~A.}\ \bibnamefont
  {Hendrickson}} \emph {et~al.},\ }\href@noop {} {\bibfield  {journal} {\bibinfo
  {journal} {Proteins: Struct., Funct., Bioinf.}\ }\textbf {\bibinfo {volume}
  {4}},\ \bibinfo {pages} {77} (\bibinfo {year} {1988})}\BibitemShut {NoStop}%
\bibitem [{\citenamefont {Hauptman}(1986)}]{Hauptman86}%
  \BibitemOpen
  \bibfield  {author} {\bibinfo {author} {\bibfnamefont {H.}~\bibnamefont
  {Hauptman}},\ }\href@noop {} {\bibfield  {journal} {\bibinfo  {journal}
  {Angew. Chem., Int. Ed.}\ }\textbf {\bibinfo {volume} {25}},\ \bibinfo
  {pages} {603} (\bibinfo {year} {1986})}\BibitemShut {NoStop}%
\bibitem [{\citenamefont {Karle}(1986)}]{Karle86}%
  \BibitemOpen
  \bibfield  {author} {\bibinfo {author} {\bibfnamefont {J.}~\bibnamefont
  {Karle}},\ }\href@noop {} {\bibfield  {journal} {\bibinfo  {journal}
  {Science}\ }\textbf {\bibinfo {volume} {232}},\ \bibinfo {pages} {837}
  (\bibinfo {year} {1986})}\BibitemShut {NoStop}%
\bibitem [{\citenamefont {Millane}(1990)}]{Millane90}%
  \BibitemOpen
  \bibfield  {author} {\bibinfo {author} {\bibfnamefont {R.~P.}\ \bibnamefont
  {Millane}},\ }\href@noop {} {\bibfield  {journal} {\bibinfo  {journal} {J.
  Opt. Soc. Am. A}\ }\textbf {\bibinfo {volume} {7}},\ \bibinfo {pages} {394}
  (\bibinfo {year} {1990})}\BibitemShut {NoStop}%
\bibitem [{\citenamefont {Bethe}\ and\ \citenamefont
  {Salpeter}(2008)}]{Bethe08}%
  \BibitemOpen
  \bibfield  {author} {\bibinfo {author} {\bibfnamefont {H.~A.}\ \bibnamefont
  {Bethe}}\ and\ \bibinfo {author} {\bibfnamefont {E.~E.}\ \bibnamefont
  {Salpeter}},\ }\href@noop {} {\emph {\bibinfo {title} {Quantum Mechanics of
  One- and Two-Electron Atoms}}}\ (\bibinfo  {publisher} {Dover},\ \bibinfo
  {address} {Mineola, NY},\ \bibinfo {year} {2008})\BibitemShut {NoStop}%
\bibitem [{\citenamefont {Ravelli}\ \emph {et~al.}(2003)\citenamefont
  {Ravelli}, \citenamefont {Leiros}, \citenamefont {Pan}, \citenamefont
  {Caffrey},\ and\ \citenamefont {McSweeney}}]{Ravelli03}%
  \BibitemOpen
  \bibfield  {author} {\bibinfo {author} {\bibfnamefont {R.~B.}\ \bibnamefont
  {Ravelli}} \emph {et~al.},\ }\href@noop
  {} {\bibfield  {journal} {\bibinfo  {journal} {Structure}\ }\textbf {\bibinfo
  {volume} {11}},\ \bibinfo {pages} {217} (\bibinfo {year} {2003})}\BibitemShut
  {NoStop}%
\bibitem [{\citenamefont {Rohringer}\ and\ \citenamefont
  {Santra}(2007)}]{Rohringer07}%
  \BibitemOpen
  \bibfield  {author} {\bibinfo {author} {\bibfnamefont {N.}~\bibnamefont
  {Rohringer}}\ and\ \bibinfo {author} {\bibfnamefont {R.}~\bibnamefont
  {Santra}},\ }\href@noop {} {\bibfield  {journal} {\bibinfo  {journal} {Phys.
  Rev. A}\ }\textbf {\bibinfo {volume} {76}},\ \bibinfo {pages} {033416}
  (\bibinfo {year} {2007})}\BibitemShut {NoStop}%
\bibitem [{\citenamefont {Persson}\ \emph {et~al.}(2001)\citenamefont
  {Persson}, \citenamefont {Lunell}, \citenamefont {Sz{\"o}ke}, \citenamefont
  {Ziaja},\ and\ \citenamefont {Hajdu}}]{Persson01}%
  \BibitemOpen
  \bibfield  {author} {\bibinfo {author} {\bibfnamefont {P.}~\bibnamefont
  {Persson}} \emph {et~al.},\ }\href@noop {} {\bibfield
  {journal} {\bibinfo  {journal} {Protein Sci.}\ }\textbf {\bibinfo {volume}
  {10}},\ \bibinfo {pages} {2480} (\bibinfo {year} {2001})}\BibitemShut
  {NoStop}%
\bibitem [{\citenamefont {Ziaja}\ \emph {et~al.}(2005)\citenamefont {Ziaja},
  \citenamefont {London},\ and\ \citenamefont {Hajdu}}]{Ziaja05}%
  \BibitemOpen
  \bibfield  {author} {\bibinfo {author} {\bibfnamefont {B.}~\bibnamefont
  {Ziaja}}, \bibinfo {author} {\bibfnamefont {R.~A.}\ \bibnamefont {London}}, \
  and\ \bibinfo {author} {\bibfnamefont {J.}~\bibnamefont {Hajdu}},\
  }\href@noop {} {\bibfield  {journal} {\bibinfo  {journal} {J. Appl. Phys.}\
  }\textbf {\bibinfo {volume} {97}},\ \bibinfo {pages} {064905} (\bibinfo
  {year} {2005})}\BibitemShut {NoStop}%
\end{thebibliography}
%


\end{document}